\documentclass[apj]{emulateapj}
\usepackage{apjfonts}
\usepackage{graphicx}
\usepackage{amsmath}
\usepackage{amssymb}
\usepackage{amsfonts}
\shorttitle{Study of CXOU J132527.6-430023}
\shortauthors{Burke et al.}

\begin{document}

\title{A Transient Sub-Eddington Black Hole X-ray Binary Candidate \\ in the Dust lanes of Centaurus A}

\author{Mark J. Burke\altaffilmark{1}, 
Somak Raychaudhury\altaffilmark{1}, 
Ralph P. Kraft\altaffilmark{2}, 
Nicola J. Brassington\altaffilmark{3}, 
Martin J. Hardcastle\altaffilmark{3}, \\
Joanna L. Goodger\altaffilmark{3},
Gregory R. Sivakoff\altaffilmark{4}, 
William R. Forman\altaffilmark{2}, 
Christine Jones\altaffilmark{2},
Kristin A. Woodley\altaffilmark{6}, \\
Stephen S. Murray\altaffilmark{5,2},
Jouni Kainulainen \altaffilmark{7},
Mark Birkinshaw\altaffilmark{8},
Judith H. Croston\altaffilmark{9},
Daniel A. Evans\altaffilmark{2},\\
Marat Gilfanov\altaffilmark{10,14}, 
Andr\'es Jord\'an\altaffilmark{2,11},
Craig L. Sarazin\altaffilmark{12},
Rasmus Voss\altaffilmark{13}, 
Diana M. Worrall\altaffilmark{8},\\ and
Zhongli Zhang\altaffilmark{10}
}

\altaffiltext{1}{School of Physics and Astronomy, University of Birmingham, Edgbaston, Birmingham, B15 2TT, UK;  \email{mburke@star.sr.bham.ac.uk}}

\altaffiltext{2}{Harvard-Smithsonian Center for Astrophysics, 60 Garden Street,Cambridge, MA 02138, USA}

\altaffiltext{3}{School of Physics, Astronomy, and Mathematics, University of Hertfordshire, Hatfield, AL10 9AB, UK}

\altaffiltext{4}{Department of Physics, University of Alberta,
Edmonton, Alberta T6G 2E1, Canada}

\altaffiltext{5}{Department of Physics and Astronomy, University of British Columbia, Vancouver BC V6T 1Z1, Canada}

\altaffiltext{6}{Department of Physics and Astronomy, Johns Hopkins University, 3400 N. Charles Street, Baltimore, MD 21218, USA}

\altaffiltext{7}{Max-Planck-Institute for Astronomy, K\"{o}nigstuhl 17, 69117 Heidelberg,Germany}

\altaffiltext{8}{HH Wills Physics Laboratory, University of Bristol, Tyndall Avenue, Bristol BS8 1TL, UK}

\altaffiltext{9}{School of Physics and Astronomy, University of Southampton, Southampton, SO17 1BJ, UK}

\altaffiltext{10}{Max Planck Institut f\"{u}r Astrophysik, Karl-Schwarzschild-Str. 1, D-85741, Garching, Germany}

\altaffiltext{11}{Departamento de Astronom\'ia y Astrof\'isica, Pontificia Universidad Cat\'olica de Chile, 7820436 Macul, Santiago, Chile}

\altaffiltext{12}{Department of Astronomy, University of Virginia, P.O. Box 400325, Charlottesville, VA 22904-4325, USA}

\altaffiltext{13}{Department of Astrophysics/IMAPP, Radboud, University Nijmegen, PO Box 9010, NL-6500 GL Nijmegen, the Netherlands.}

\altaffiltext{14}{Space Research Institute, Russian Academy of Sciences, Profsoyuznaya 84/32, 117997 Moscow, Russia}

\begin{abstract}
We report the discovery of a bright X-ray transient, CXOU J132527.6-430023, in the nearby early-type galaxy NGC~5128. The source was first detected over the course of five \emph{Chandra} observations in 2007, reaching an unabsorbed outburst luminosity of ${1-2\times 10^{38}}$~${\rm erg~s^{-1}}$ in the 0.5-7.0 keV band before returning to quiescence.  Such luminosities are possible for both stellar-mass black hole and neutron star X-ray binary transients. Here, we attempt to characterize the nature of the compact object. No counterpart has been detected in the optical or radio sky, but the proximity of the source to the dust lanes allows for the possibility of an obscured companion.  The brightness of the source after a $>$100 fold increase in X-ray flux makes it either the first confirmed transient non-ULX black hole system in outburst to be subject to detailed spectral modeling outside the Local Group, or a bright ($>10^{38}~{ \rm erg~s^{-1}}$) transient neutron star X-ray binary, which are very rare. Such a large increase in flux would appear to lend weight to the view that this is a black hole transient. X-ray spectral fitting of an absorbed power law yielded unphysical photon indices, while the parameters of the best-fit absorbed disc blackbody model are typical of an accreting ${\sim 10~M_\odot}$ black hole in the thermally dominant state.      
\end{abstract}


\section{Introduction}
\label{s:intro}
The majority of confirmed black hole low- and high-mass X-ray binary (BH LMXB, HMXB) systems are transient sources that experience outbursts lasting from a few weeks to a matter of years.  These outbursts are characterized by an increase in X-ray luminosity by many orders of magnitude from a quiescent state of $L_X\sim10^{30-34}~{\rm erg~s^{-1}}$, recurring on timescales of months to decades (for a review, see Remillard \& McClintock 2006).  The recurrent behavior of the persistent luminosity, characterised by a rapid rise and exponential decay, is thought to be the result of accretion disc instability (Dubus, Hameury, \& Lasota 2001).  Research into Galactic BH LMXBs has revealed three common spectral states, between which sources are observed to transit; a thermal state dominated by radiation originating in the inner-regions of a geometrically thin, optically thick accretion disk (Shakura \& Sunyaev 1973), a hard state dominated by a power law with photon index $\Gamma\sim1.7$, and a steep power law state $\Gamma\sim2.5$ extending to MeV energies where a significant thermal component is also present. The similarity of these states to the spectra of neutron star (NS) LMXB means that it is often impossible to fully distinguish between bright NS and BH LMXBs on the basis of luminosity and spectra alone.  However, a distinction can be drawn between the thermal states, which are typically much harder in NS systems than for BH sources.  This is in agreement with the proposal that `ultrasoft' spectra are a characteristic of BH sources (White \& Marshall, 1984).  NS systems, as a result of the presence of the physical surface of the compact object,  experience behavior in the form of bursting, for lower-luminosity (Atoll) sources such as 4U 1636-536, and flaring, for more luminous ($L_X \sim 10^{38}~{\rm erg~s^{-1}}$, Z-track) systems such as Sco X-1.  

Until recently only two transient NS LMXBs with $L_X\sim L_{Edd}$ were known, Cir X-1 and XTE J1806-246.  Cir X-1 is a peculiar NS XB with atypical spectral and timing behavior and was once thought to possess a BH primary, an idea which was confounded by the detection of Type-I X-ray bursts (Tennant et al. 1986, Linares et al. 2010).  The transient nature of Cir X-1 is also unusual in that it is periodic with a period of 16.6 days and therefore thought to be the result of the eccentric orbit of the donor star (Murdin et al. 1980), rather than accretion disc instability.  The peak luminosity of XTE J1806-246, reported as $L_X\sim1.5\times10^{38}~{\rm erg~s^{-1}}$ (Wijnands \& van der Klis 1999), is based on a poor distance estimate (8 kpc, based on a sky position coincident with the Galactic bulge).
The recent discovery of transient NS LMXB, XTE J1701-462 (Homan et al. 2007), which experienced the full range of spectral and timing behavior observed in all other Galactic NS LMXBs (Lin, Remillard \& Homan 2009, Sanna et al. 2010) during a $\sim$ 600 day outburst, demonstrates both that NS transients can be as luminous as some of their BH cousins, and that the short-term variability is fundamentally linked to the accretion rate of the system.  
\begin{figure*}[htb!]\center
{\includegraphics[width=0.8\hsize]{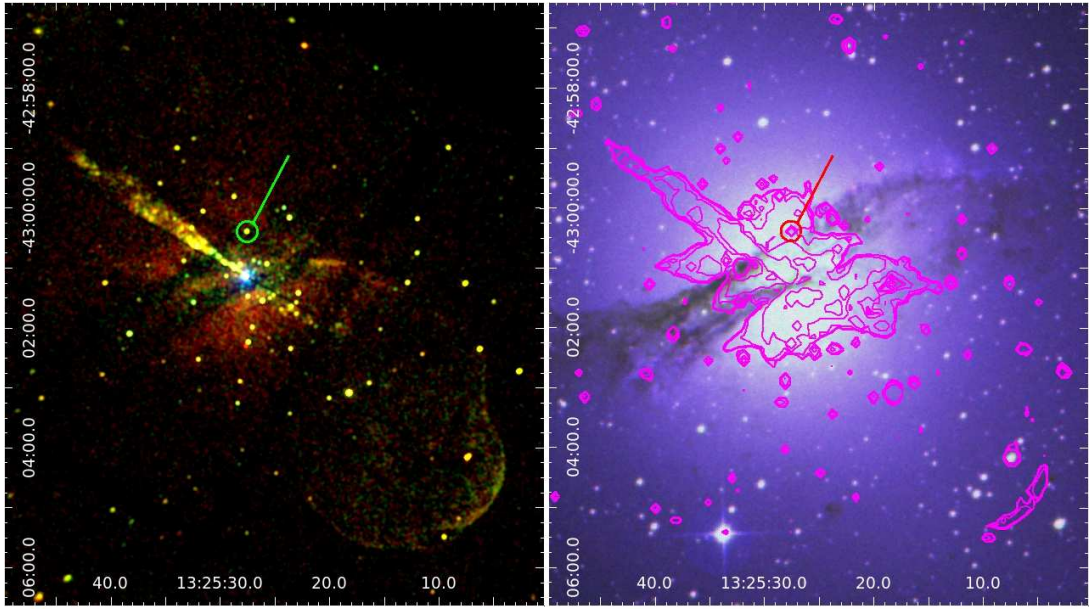}}
\caption{Location of CXOU J132527.6-430023 within Cen~A.  \emph{Left:} A Gaussian-smoothed logarithmic-scale 0.5-1.3 keV (red), 1.3-2 keV (green) 2-7 keV (blue) image of ObsID 7797.  \emph{Right:} X-ray 0.5-7 keV contours of the brighter central regions of Cen~A (Magenta) overlaid on Deep Sky Survey combined R, I and B band image of Cen~A.
\label{fig:pos}}
\end{figure*}

The \emph{ Chandra} and \emph{ XMM-Newton} X-ray observatories have made possible the detailed study of extragalactic X-ray binary (XB) sources during the last decade. Within the Local Group, Williams et al. (2006) cataloged the properties of 45 transient XBs in M 31, and found that most exist around the central bulge of the galaxy, suggesting that they are part of an older stellar population and therefore more likely to be LMXBs. More detailed studies of brighter transient sources (Trudolyubov et al. 2006) and the overall XB population (Stiele et al. 2011) show many soft sources, with absorbed disc blackbody temperatures $kT_{in}< 1$~keV, and $<$ 0.1 keV in the case of super-soft sources (SSS) that show no emission above 1 keV.  Such low values of $kT_{in}$ are inconsistent with spectral fitting results for Galactic NS LMXB (e.g. Revnivtsev \& Gilfanov 2006) and are thus indicative of BH systems in the thermal state, while SSS are thought to be accreting white dwarfs (van den Heuvel et al. 1992, Kahabka \& van den Heuvel 1997).  It has been shown that the M 31 X-ray transients are found in three distinct spectral groups at their peak luminosity (Voss et al. 2008), categorized as very soft, soft and hard, half of all the transients being hard.  The ChASeM33 project, consisting of ${  7\times200~{\rm ks}}$ \emph{ Chandra} exposures of M 33, allowed source detection to a depth of ${5\times10^{34}}~{\rm ergs~s^{-1}}$, and 7 transient sources have been discovered, albeit only two with $  L_X > 10^{37}~{\rm erg~s^{-1}}$ (Williams et al. 2008).   

Outside the Local Group, in the elliptical galaxies NGC 3379 and NGC 4278 (Brassington et al. 2008, 2009), five and three sources respectively have been classified as transient candidates (TCs).  Three other sources in each of these galaxies have been identified as possible transient candidates (PTCs).  TCs and PTCs correspond to observed changes in luminosity by factors of $>10$ and $>5$ respectively, and we note that some Galactic sources such as 4U 1705-44 (Homan et al. 2009) and 4U 0513-40 (Maccarone et al. 2010) are not transients but do show variations significantly $>10$.  All of these sources have hardness ratios and X-ray colors that are consistent with Galactic LMXBs.  Follow-up spectral analysis on the bright, seemingly persistent sources in these galaxies (Brassington et al. 2010, Fabbiano et al. 2010) yielded $L_X \sim 10^{38}-10^{39} {~\rm erg~s^{-1}}$, with spectral fitting parameters congruent with known BH systems; the strongest BH binary candidate had maximum and minimum unabsorbed luminosities of $1.2\times10^{39} {~\rm erg~s^{-1}}$ and $4.1\times10^{38} {~\rm erg~s^{-1}}$.  

For observational reasons, the XBs outside the Local Group that have been the most intensely studied are the Ultraluminous X-ray sources (ULXs), which possess outburst luminosities in excess of $10^{39}~\rm{erg~s^{-1}}$, together with other bright sources with luminosities typically $>2$--$3 L_{Edd}$ for an accreting NS.  These sources represent a comparatively small fraction of XBs in the Local Group, with the solitary ULXs found in M 31 and M 33 being the only constituents of their class (Kaur et al 2011, Dubus \& Rutledge 2002).  Therefore, there exists an asymmetry between the study of XBs inside and outside the Local Group.  
\begin{deluxetable}{lcccr}
\tabletypesize{\footnotesize}
\tablewidth{0pt}
\tablecaption{ACIS-I/S Observations of Transient in Cen A \label{tab:trans2}}
\tablehead{
\colhead{ObsID} &
\colhead{Instrument} &
\colhead{Exposure} &
\colhead{Date} &
\colhead{Net Counts\tablenotemark{a}} \\
\colhead{} &
\colhead{} &
\colhead{(ks)} &
\colhead{MJD~~~(y-m-d)} &
\colhead{}
}
\tablecolumns{4}
\startdata
316 &	ACIS-I & 36.18 & 51517~~~(1999-12-05) & $<2.7$ \\
962 &	ACIS-I & 36.97 & 51681~~~(2000-05-17) & $<4.3$ \\
2978 & ACIS-S & 45.18  & 52520~~~(2002-09-03) & $<4.3$ \\
3965 & ACIS-S & 50.17 & 52896~~~(2003-09-14) & $<4.0$ \\
7797 &  ACIS-I & 	98.17 & 54181~~~(2007-03-22)  & $\mathbf{336\pm32}$ \\
7798 &  ACIS-I & 	92.04 & 54186~~~(2007-03-27) & $\mathbf{364\pm34}$ \\
7799 &  ACIS-I & 	96.04 & 54189~~~(2007-03-30) & $\mathbf{375\pm35}$ \\
7800 &  ACIS-I & 	92.05 & 54207~~~(2007-04-17) & $\mathbf{284\pm31}$ \\
8489 &  ACIS-I & 	95.18 & 54228~~~(2007-05-08)  & $\mathbf{16^{+9}_{-7}}$ \\
8490 &  ACIS-I & 	95.68 & 54250~~~(2007-05-30)  & $<7.3$ \\
10723 & 	ACIS-I & 5.15  & 54835~~~(2009-01-04)  & $<2.6$ \\
10724 & 	ACIS-I & 5.17  & 54895~~~(2009-03-05)  & $<2.6$ \\
10725 & 	ACIS-I & 5.04  & 54947~~~(2009-04-26)  & $<2.6$ \\
10726	 & ACIS-I & 5.15 & 55003~~~(2009-06-21)  & $<4.0$ \\
10722 & 	ACIS-S & 50.04 & 55082~~~(2009-09-08)  & $<2.3$ \\
11846	 & ACIS-I & 4.75  & 55312~~~(2010-04-26)  & $<3.9$ \\
11847 & 	ACIS-I & 5.05  & 55455~~~(2010-09-16)  & $<2.6$ \\
12155 & 	ACIS-I & 5.05  & 55552~~~(2010-12-22)  & $<2.6$ \\
12156 & 	ACIS-I & 5.06  & 55734~~~(2011-06-22) & - \\
\enddata
\tablenotetext{a}{For observations where the source was not detected, the Net Counts column represents the 90\% upper limit placed on the net counts of the source in the 0.5-2.0 keV band. For the {\bf detections}, 90\% confidence regions are shown.  The position of the source is coincident with the ACIS readout-streak in ObsID 12156 and no upper-limit was calculated in this instance.}
\end{deluxetable}

\subsection{NGC 5128}
Centaurus A (NGC 5128, Cen~A) is the nearest optically luminous large early-type galaxy, situated at a distance of 3.7 Mpc (Ferrarese et al. 2007), with $M_B$ = -21.1. A small late-type galaxy is currently merging  with Cen~A,  however, the galaxies remain poorly mixed (Quillen et al. 2006).  The merger has resulted in the presence of vast dust lanes  that obscure many of the central regions at optical and soft X-ray wavelengths (Graham 1979).   
\begin{figure}[htb!]\center\rotatebox{270}
{\includegraphics[height=3.5in]{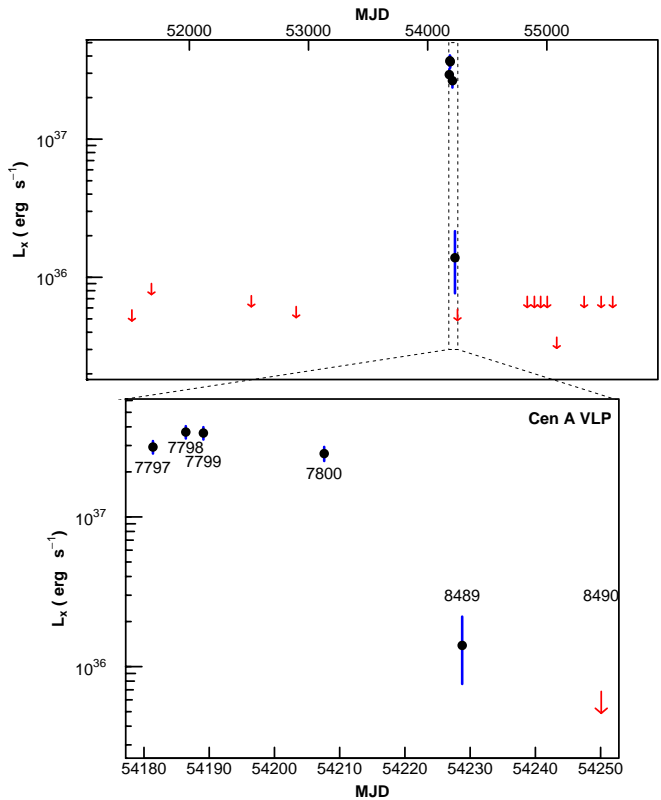}}
\caption{X-ray lightcurve (0.5 -- 2 keV),  uncorrected for absorption.  \emph{Top:} Luminosity is displayed with 90\% confidence intervals (blue error bars) and with 90\% upper limits from other ACIS observations (red arrows).  A combined upper limit was obtained from the 5 ks observations (the seven arrows to the right).  \emph{Bottom:} Lightcurve of VLP observations. 
\label{fig:LC}}
\end{figure}

Six 100~ks \emph{ Chandra} observations were taken as part of the Cen~A Very Large Project (VLP) spanning the course of 2 months (Jord{\'a}n et al. 2007).  These observations have allowed unprecedented insight into the X-ray jets (Hardcastle et al. 2007, Worrall et al. 2008, Goodger et al. 2010), radio-lobe shock (Croston et al. 2009), extended gaseous emission (Kraft et al. 2008) and work on the XB, particularly in relation to globular clusters (Voss et al. 2009, see also Voss \& Gilfanov 2006 and Woodley et al. 2008 using pre-VLP data).

Two transients in the field of Cen~A have previously been studied.  Sivakoff et al. (2007) discovered a previously undetected transient using VLP observations, CXOU J132518.2-430304. With outburst luminosities in excess of ${10^{39}}~{\rm erg~s^{-1}}$, it is by definition an ULX.  CXOU J132518.2-430304 underwent a >770-fold increase in luminosity between quiescent and active states, with an outburst duration of >70 days. Spectral analysis coupled with \emph{ HST} observations strongly suggest that the source is a BH LMXB that transitions from the steep power law state to the thermally dominant state over the course of the VLP observations.  Prior to this, Kraft et al. (2001) had detected a transient ULX,  CXOU J132519.9-43031, ($  L_X\sim 10^{39}~{\rm erg~s^{-1}}$) within ${  \sim 6\arcsec}$ of a source detected by \emph{ROSAT} over the course of 10 days in 1995.
Spectral fitting suggested the source was in a steep power law state (see also Ghosh et al. 2006).  

In this paper, we present the first detailed analysis of a sub-Eddington X-ray transient at Mpc distances, CXOU J132527.6-430023, whose flux rose by more than a factor of $100$ from quiescence to outburst and reached peak unabsorbed luminosities of $\sim2\times10^{38}~{\rm erg~s^{-1}}$ (0.5-7.0 keV) before fading back to a quiescent state by MJD 54250 (Table~\ref{tab:trans2}).  We present evidence that this source is a BH LMXB and compare it to LMXB systems observed both inside and outside the Local Group.  This source is clearly a bridge between more typical Galactic and extragalactic XBs. Since Cen~A is nearby, this is the only sub- (or $\sim$) Eddington transient where we can perform detailed spectral fitting and attempt to classify the compact as a BH or NS.

\section{Data Preparation and Analysis}
\label{s:obs}
\subsection{Preliminary Work}

An overview of all \emph{Chandra} ACIS observations of Cen~A can be found in Table~\ref{tab:trans2}.  All of the 5 ks and 50 ks observations of Cen~A were taken with the Advanced CCD Imaging Spectrometer (ACIS) as part of the HRC Guaranteed Observation Time program (PI: Murray).

All the data were reprocessed using CIAO 4.3 and HEASOFT 6.11 with CALDB 4.4.2.  All observations were first co-aligned to a precision $<0\farcs2$ based on the positions of the point sources detected using {\tt wavdetect} and then aligned to the positions of confirmed Cen~A globular clusters.  This procedure will be fully described in a later paper.  CXOU J132527.6-430023 was detected in \emph{Chandra} ObsIDs 7797, 7798, 7799, 7800, and 8489. 
 \begin{figure*}[htb!]\center\rotatebox{270}
{\includegraphics[width=2in]{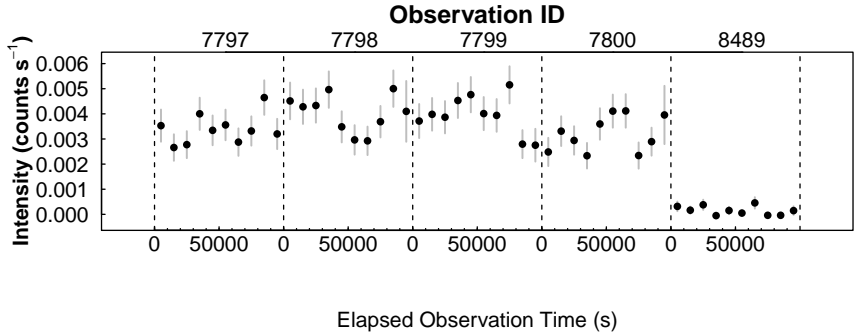}}
\caption{Intra-Observation 0.5 -- 2.0 keV lightcurves of the net intensity with 10 ks binning within the VLP observations.  For the ObsID dates, see Table~\ref{tab:trans2}.
\label{fig:LC2}}
\end{figure*}

We measured the position of CXOU J132527.6-430023 as ${ \alpha =13^{h}25^m27.58^s}$, $\delta=-43^\circ 00\arcmin 23\farcs3$ (J2000) with $<0\farcs2$ uncertainty (Fig.~\ref{fig:pos}). The 0.5-2.0 keV lightcurve across the 100 ks observations was generated using the ciao tool {\tt aprates}.  The source flux was calculated based on a circular region of radius $r_{p2}$ equal to 90\% of the 2 keV PSF centred at the off-axis position of the source, and the background flux from an annulus extending from $r_{p2}$ to $3r_{p2}$.  No other sources are seen to overlap these source regions, and no other $r_{p2}$ source regions encroach on the background annuli. In Fig.~\ref{fig:LC}, we present lightcurves in the 0.5-2.0 keV band, with the VLP  (outburst) lightcurve shown in the bottom panel, while the top panel shows the overall ACIS lightcurve of the source.  We restricted ourselves to this low-energy range due to the proximity of the source to the AGN. For the six 100 ks observations, the mean photon energies per effective exposure of each 0.5-2.0 keV event were first calculated using {\tt eff2evt} on both source and background regions and are then used as {\tt aprates} input parameters.  For the upper limits, the exposure times of the source and background regions were also used.  {\tt aprates} then performs a Bayesian analysis and returns the net 0.5-2.0 keV flux with a 90\% confidence interval or count-rate upper limit.  A combined upper limit was calculated for the $7\times5$ ks observations and individual upper limits were obtained for each observation of exposure ${  > 30~{\rm ks}}$.  The upper limits were then converted to 0.5-2.0 keV fluxes, uncorrected for absorption, using the Chandra X-Ray Center program {\tt PIMMS}, assuming an absorbed power law spectrum with $\Gamma=2.0$ and $N_H=0.084\times10^{22}~{\rm cm^{-2}}$ (i.e. a relatively steep spectrum for a quiescent LMXB in front of the dust lane).      The source position is obscured by the ACIS readout streak in ObsID 12156, no upper limit was obtained in this case.  With {\tt aprates}, the combined upper limit on count rate was calculated by inputting the summation of counts, areas and exposure times of the source and background regions. Measured fluxes and upper limits were then converted into luminosity using a distance of 3.7 Mpc.  We recognize that a more recent measurement has produced a distance to Cen~A of $3.8\pm0.1$ Mpc (Harris et al. 2010), but we have used a distance of 3.7 Mpc in calculations throughout this paper for the sake of consistency with earlier work.  This said, the uncertainty of the distance measurement is much less than the uncertainties associated with the calibration and spectral fitting process, which therefore have a greater effect on our luminosity measurement.  ObsID 10722 is our lowest luminosity upper limit, $L_X > 3\times10^{35}{\rm ~erg~s^{-1}}$.

Several factors make CXOU J132527.6-430023 the best transient BH candidate in Cen~A to study in the $\sim10^{38} {\rm ~erg~s^{-1}}$ regime.  It is clear that it is observed in both outburst and quiescent states during the VLP observations.  Basic observational criteria regarding the source are satisfied; the source is always in the field of view, far from the ACIS chip edges and has the required counts to perform meaningful spectral analysis.  Our work, currently in preparation, shows that no other source, observed to be transient in Cen~A during the VLP, is as bright in as many observations while also undergoing such a dramatic change in the mean flux (Fig. \ref{fig:LC}), as CXOU J132527.6-430023.

Before the extraction of spectra, we investigated the intra-observation variability, by extracting a count-rate lightcurve of each observation, to test if the source was active throughout a given observation, or was prone to more discreet behavior.  In Fig. \ref{fig:LC2} we present the intra-observation 0.5-2.0 keV lightcurve, with count-rate grouped to 10 ks bins.  The mean intensity has substantially diminished by the time of ObsID 8489, approximately 20 days after the previous observation.  A simple $\chi^2$ test showed no significant variations from the best-fit straight line for any of the 100 ks \emph{ Chandra} pointings binned to 10 ks.

\subsection{Possible Counterparts}
\label{sec:multi}
We investigated the possibility of detecting a counterpart to CXOU J132527.6-430023 at other wavelengths.  We examined B and R band images from recent observations (Harris et al. 2011) with the Walter Baade Telescope and the Inamori Magellan Areal Camera (IMACS) and found no point sources coincident with the position of the transient.  Further investigation using an \emph{HST} WFPC2  observation, U4100108M (F814W) showed no optical counterpart.  However, the position is coincident with the edge of the dust lane.  This being the case, any attempt to perform optical photometry on the source position will be inconclusive.  The nearest globular cluster that has been spectroscopically confirmed as belonging to Cen~A is ${\sim  30\arcsec}$ away from CXOU J132527.6-430023 (Woodley et al. 2007).  However, no clusters have been detected within the radial distance of the transient from the galaxy center, as a result of the high stellar luminosity within $\sim1\farcm5$ of the Cen A nucleus, in tandem with the heavy obscuration by the dust lanes.

A 2007 June Very Large Array 8.4~GHz  observation using the A configuration, obtained just one week after the final VLP observation, places a $3\sigma$ upper limit of 0.13 mJy on the radio emission at this position.  Further to this, it has been confirmed that no radio counterpart has been detected in any other VLP observation (Goodger et al. 2010). Galactic transient XBs frequently show evidence of jets during outburst (see Fender 2005 for review), and have a clear relationship between the strength of the outburst X-ray luminosity and the flux density of the radio emission.  Sources typically possess a flat radio spectrum, with flux densities $\sim1-1000$ mJy when scaled to 1 kpc.  Scaling this to the distance to Cen~A,  suggests that the expected flux density of a typical counterpart would lie in the range $0.07 {\rm n Jy}$ to $0.07 {\rm \mu Jy}$, far smaller than our upper limit.  Cyg X-3 is a HMXB where the evidence of a BH is fairly strong (Shrader et al. 2010) and has been seen to emit strong radio flares of up to 20 Jy from an assumed distance of 9 kpc (Corbel et al. 2012); at the distance of Cen A such an outburst would be detected above our upper limit.  Therefore a radio detection of an XB counterpart in Cen A would point towards a high mass companion.

\begin{figure}[htb!]\center
{\includegraphics[width=3.5in]{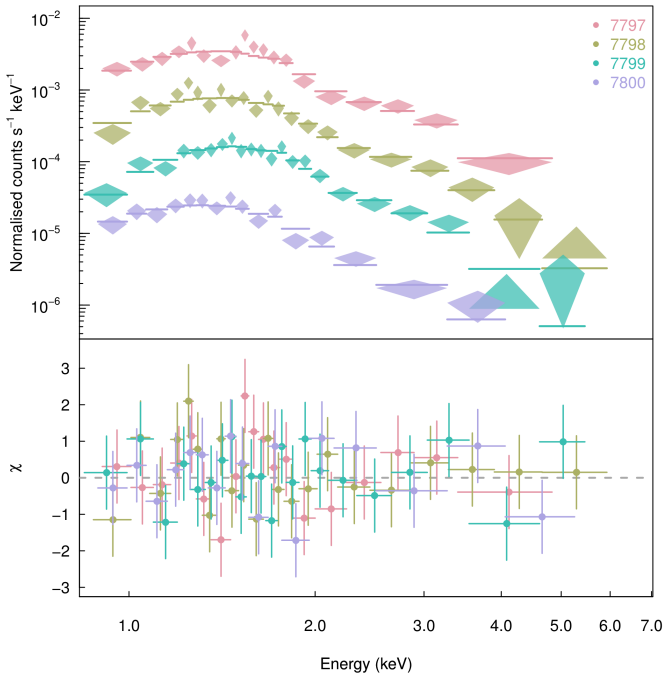}}
\caption{Spectra of the four bright VLP observations of CXOU J132527.6-430023.  Spectra subsequent to 7797 are offset in $y$ by a factor ${  \times5^{-n}}$, where n=1 for 7798, n=2 for 7799 etc.  Spectra values and associated errors are represented by diamonds while the absorbed disc blackbody model is represented by horizontal lines.  Triangles show instances where the uncertainty was greater than the rate in the channel. 
\label{fig:spec}}
\end{figure}

\ref{fig:spec}).  
\subsection{Spectral Fitting}
Spectra and responses in the 0.5-7.0 keV range for each of the five observations where the source is detected were generated, based on circular apertures of radius $r_{p7}$ equal to 90\% of the 7 keV PSF size at the off-axis source position. Background spectra were taken from an annulus $2r_{p7}$ to $4r_{p7}$, minus two overlapping circular $r_{p2}$ source regions corresponding to two faint sources that were only detected in the merged 600 ks dataset.  In the four bright observations, 7797 -- 7800, the spectra were grouped to a minimum of 20 counts $\mathrm{channel^{-1}}$ to allow use of the $\chi^2$-statistic, whereas the spectrum from observation 8489 contained too few counts for grouping and fitting.  All fitting was performed using XSPEC 12.7.0 (Arnaud, 1996).

While the large range in flux from the source negates the possibility of it being a background object, such as an AGN, it is prudent to check that the source is not a foreground flare star.  We tested this by attempting to fit an absorbed thermal plasma model ($phabs \times apec$).  For the default settings (abundance fixed at 1.0) fit statistics for the spectra from ObsId 7797 to ObsId 7800 were $\chi^2/{\rm dof} =56.52/17, 58.1/20, 102.4/20$ and $44.1/15$  (${\chi^2_\nu  > 2}$), respectively.  Allowing the abundance to vary led to good fits, ${  \chi^2_\nu \sim 0.8-1.4}$, but the abundance tended towards zero, approximating a pure bremsstrahlung spectrum. Coupled with an apparent lack of optical counterpart it is reasonable to discount the possibility of CXOU J132527.6-430023 being a foreground object.

We followed the spectral fitting prescription of Brassington et al. (2010), who developed a consistent approach to fitting the spectra of luminous LMXBs in NGC 3379.  Absorbed single-component models, starting with a simple power law, are fit first, and the nature of these fits informs a decision on how to proceed to further fitting.  Each observation was individually fit in XSPEC with a $phabs\times powerlaw$ model, all parameters of which were left free.  Successful fits were achieved for the four observations where the source was bright, with the largest ${  \chi^2_\nu}\sim 1.1$.  The fitting yielded quite high values for the photon index, $\Gamma\sim 3.7-4.7$, while the absorption column was above the Galactic value of ${  N_H=0.084  \times10^{22}~{\rm cm^{-2}}}$ (Dickey \& Lockman 1990), with ${  N_H \sim 0.9-1.7  \times10^{22}~{\rm cm^{-2}}}$, also larger than the inferred range of the dust lanes and in the $2\arcsec$ vicinity of the source (\S~\ref{sec:multi}). The results of the simulations of Brassington et al. (2010)  suggest that XSPEC absorbed power law fits on LMXB will increase the value of ${N_H}$ as a compensation for instances where there is a lack of a required thermal component.  It was therefore appropriate to fit an absorbed multicolored disc blackbody model ($phabs \times  diskbb$ in XSPEC) and contrast the two sets of fit results. 

\begin{deluxetable}{cccccc}
\tabletypesize{\footnotesize}
\tablewidth{0pt}
\tablecaption{Best Fit Parameter Values: $phabs \times diskbb$ Model\label{tab:trans1}}
\tablehead{
\colhead{ObsID} &
\colhead{$\mathrm{N_H}$} &
\colhead{Flux\tablenotemark{a} \tablenotemark{b}} &
\colhead{$\mathrm{kT_{in}}$} &
\colhead{$\mathrm{\chi^2/dof}$} &
\colhead{$\mathrm{L_{x}}$\tablenotemark{a} \tablenotemark{b} \tablenotemark{c}} \\
& 
\colhead{${ \rm 10^{22}~cm^{-2}}$} & 
\colhead{${ \rm 10^{-14}}$ \tablenotemark{d}} &
\colhead{keV} &
& 
\colhead{$\mathrm{10^{38}~erg~s^{-1}}$}
}
\tablecolumns{6}
\startdata
7797 & $0.36_{-0.19}^{+0.22}$ & $7.69_{-1.44}^{+2.04}$ & $0.65_{-0.10}^{+0.11}$ & 15.85/16  & $1.26_{-0.24}^{+0.33}$\\
7798 & $0.43_{-0.18}^{+0.22}$ & $7.93_{-1.49}^{+2.22}$ & $0.61_{-0.10}^{+0.12}$ & 15.18/19 & $1.30_{-0.25}^{+0.36}$\\
7799 & $0.83_{-0.23}^{+0.29}$ & $12.12_{-2.97}^{+5.05}$ & $0.51_{-0.08}^{+0.09}$ & 11.83/19 & $1.99_{-0.49}^{+0.82}$\\
7800 & $0.33_{-0.22}^{+0.27}$ & $4.30_{-1.07}^{+1.92}$ & $0.53_{-0.11}^{+0.13}$ & 11.75/14 & $0.70_{-0.17}^{+0.32}$\\ 
\enddata
\tablenotetext{a}{Spectral fits to the 0.5 -- 7.0 keV band.}
\tablenotetext{b}{Unabsorbed, from the XSPEC $cflux$ parameter.}
\tablenotetext{c}{$L_X$ based on distance of 3.7 Mpc.}
\tablenotetext{d}{${\rm ergs~cm^{-2}~s^{-1}}$}
\end{deluxetable}

The absorbed disc blackbody was the best-fit model for all four of the bright spectra (Table \ref{tab:trans1}), with ${  \chi^2_\nu}<1$.  The addition of a second additive component, such as a power law, cut-off power law or $compTT$ proved unnecessary, as their normalizations tended towards zero.  Table \ref{tab:trans1} shows that the ${  N_H}$ in the $phabs \times diskbb$ fits is significantly less than for the $phabs\times powerlaw$ model, above the Galactic value and, for three of the spectra, consistent with the range of $N_H\sim{ (0.11-0.66)\times10^{22}~{\rm cm^{-2}}}$ inferred from Schreier et al. (1996).  However, this is still less than our calculated average value ($N_H\sim 0.6 \times 10^{22}~{\rm cm^{-2}}$) over a $2\arcsec$ region from a K-band optical depth map.  These results suggest the source exhibited minor spectral evolution across the four bright observations (Fig.~

In order to test whether the apparent difference in absorption between ObsID 7799 and the other spectra was a real effect or the result of stochastic spectral variation, a joint fit using the spectra from ObsIDs 7797, 7798 and 7800 was performed, achieving ${\chi^2_\nu} \sim 48.5/53$.  This produced more constrained parameter values of $N_H=0.38 \pm 0.12 {\times10^{22}~\rm{cm^{-2}}}$ and $kT_{in}=0.60 \pm 0.05$ keV.  The three observations would have luminosities of $1.28 \times 10^{38}~{\rm erg~s^{-1}}$, $1.22\times 10^{38}~{\rm erg~s^{-1}}$ and  $7.2 \times 10^{37}~{\rm erg~s^{-1}}$ respectively. We investigated the spectral fitting behavior in the $kT_{in}-N_H$ plane across a $50\times50$ grid of the parameter space (Fig.~\ref{fig:cont}).  These results suggest that we can reject consistency of $N_H$ at the $2\sigma$ level between the joint fit and the fit to ObsID 7799.  This does suggest that there is a genuine change in absorption local to the source during ObsID 7799.  A change in local absorption has previously been established in other BH systems.  Oosterbroek et al. (1997) simulated spectra of various model parameters in an attempt to infer which change in parameter space could reproduce the observed color-color diagrams of GS 2023+338.  They concluded that a large increase in the local absorption was the best explanation of the observed color-color tracks, and speculated that the system could be inclined enough for the absorbing material to originate in the edge of the accretion disc.  Such a change in absorption has also been reported from an extragalactic BH source, XMMU 122939.7+07533 (Maccarone et al. 2007).  This was at first attributed to a warping of the accretion disc (Shih et al. 2008), but the presence of strong, broad OIII lines (Zepf et al. 2008) suggests a strong disc wind varies the amount of absorbing material along the line-of-sight.

\section{Discussion}
\label{s:disc}
In this section we compare the properties of the transient source CXOU J132527.6-430023 to other XBs, both Galactic and extragalactic in origin, in the hope of finding well-studied analogues.  The outburst duration is between $\sim$50 and $\sim$1500 days.  This is consistent with outbursts observed in many Galactic sources, which are active on timescales of months to years -- much longer than so-called fast X-ray transients, for which the outburst duration is less than a day (Heise \& Zand 2005).  The peak unabsorbed luminosity of ${   \sim 2 \times 10^{38}}~{\rm erg~s^{-1}}$ is close to the Eddington luminosity of an accreting ${ 1.4~M_\odot}$ NS; however, it is possible that the source reached its peak luminosity prior to first being detected.  The nature of the compact object is ambiguous based solely on the luminosity.

\begin{figure}[htb!]\center\rotatebox{270}
{\includegraphics[width=3.5in]{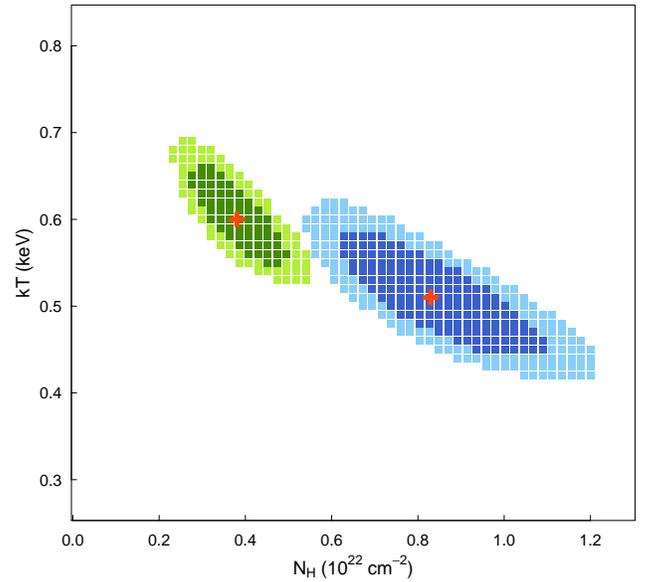}}
\caption{90\% and 68\% confidence regions with best-fit values (red crosses).  \emph{Left:} From joint spectral fit to ObsIDs 7797, 7798 \& 7800. \emph{Right:} From spectral fit of ObsID 7799.
\label{fig:cont}}
\end{figure}

\begin{figure}[htb!]\center\rotatebox{270}
{\includegraphics[height=3.5in]{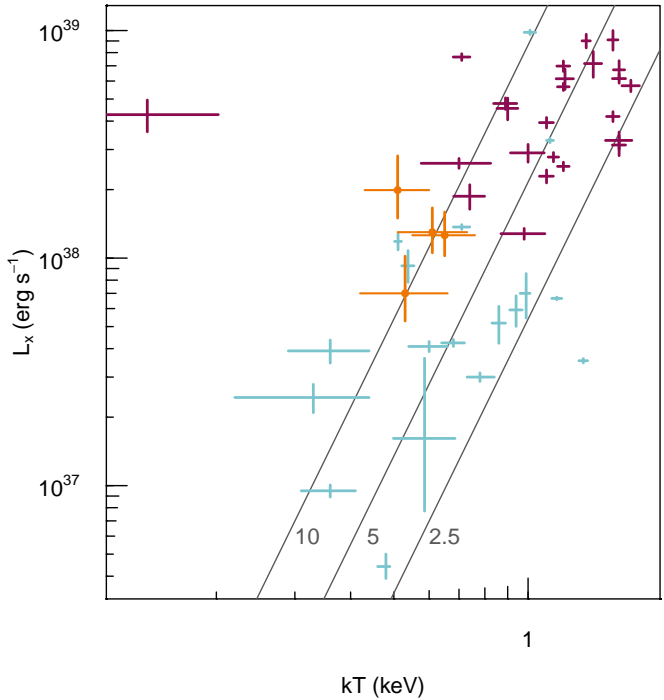}}
\caption{
\label{fig:lxkt} Comparison of $L_X-kT_{in}$ for CXOU J132527.6-430023 (orange with points) with other sources in the thermally dominant state both outside (magenta) and inside (light blue) the Local Group.  We make use of the spectral fitting results from sources in NGC 3379 (Brassington et al. 2010) and NGC 4278 (Fabbiano et al. 2010).  The Local Group sample is based on spectral fits to Galactic (McClintock and Remillard 2005, Soria et al. 2011) and M 31 (Trudolyubov 2006, Stiele et al. 2011) sources with $L_X$ adjusted to the 0.5-7.0 keV band using Xspec.  Lines of constant BH mass with increasing ${  kT_{in}}$ for 10~${  M_\odot}$, 5~${  M_\odot}$ and 2.5~${  M_\odot}$ BH are based on the prescription of Gierli{\'n}ski \& Done (2004) and assumptions of Brassington et al. (2010).}
\end{figure}

Samples of luminous LMXBs from NGC 3379 and NGC 4278 have been used to give an indication of BH mass (Brassington et al. 2010, Fabbiano et al. 2010) based on the $L_X-kT_{in}$ relation of Gierli{\'n}ski \& Done (2004).  In agreement with the broad trend shown by these more luminous sources, the spectral fitting results for three of the CXOU J132527.6-430023 spectra are compatible with a ${   \sim10~M_\odot}$ BH (Fig.~\ref{fig:lxkt}). The best-fit to ObsID 7799 yields a value of ${  N_H}$ that is double that obtained from the other spectra (albeit they are constrained to 90\% uncertainties of $\pm 50\%$).  If this is not a real effect, then it could result in an artificial increase in the unabsorbed flux found from fitting.  We find a $2\sigma$ difference between the fits for this ObsID 7799 and a joint fit of the three other bright spectra (Fig. \ref{fig:cont}), indicating that a physical change near the source has occurred.  The fitting parameters from ObsID 7799 would be consistent with an ${  \sim18~M_\odot}$ BH; however, it seems very improbable that the compact object gained, and then lost, so much mass during this time.

The spectral fitting process resulted in a best-fit model of an absorbed multicolor disc blackbody (Table \ref{tab:trans1}); the fits to the power law model lead to extremely steep spectra with $\Gamma > 3.7$ and can be ruled out.  The absorption column parameter is still above the Galactic value (${  0.084  \times10^{22}~\rm{cm^{-2}}}$); however, judging from the position of the source within Cen~A (${ \alpha =13^{h}25^m27.58^s}$, $\delta=-43^\circ 00\arcmin 23\farcs3$,~Fig.~\ref{fig:pos}), this is not unexpected. The structure of the dust lane is complex, with the source residing in an $\sim2\arcsec$ radius quasi-circular region of lower density dust bordered by regions of higher density, which complicates accurately computing the amount of absorption at the source location. All values of ${N_H}$ from the spectral fitting are consistent with studies of the extinction in the Cen~A dust lanes (Schreier et al. 1996), with ${  0.5 < A_V < 3.0}$ indicating an ${  N_H}$ of ${  (0.11-0.66)\times10^{22}~{\rm cm^{-2}}}$ (G{\"u}ver \& {\"O}zel 2009). Analysis of the K-band optical depth map presented in Kainulainen et al. (2009) yielded a mean $A_K=0.325$, indicating $N_H\sim 0.6 \times 10^{22}~{\rm cm^{-2}}$, this would be almost double the absorption column found from three of the spectral fits.   The Cen~A ULX found by Sivakoff et al. (2008) - outside of the dust lanes - was found to have a similar absorption column to CXOU J132527.6-430023 but only when the source was in the steep power law state, the column apparently decreasing to consistency with the Galactic value as the thermal component began to dominate the spectrum.  

The inner-disc temperatures, ${  kT_{in}\sim0.6~\pm0.1}$~keV, are softer than the spectra of NS LMXBs, which typically vary between 1-2 keV. This temperature is also below that of the ULX, which had ${   kT_{in} \sim1}$~keV.   Such soft spectra are reminiscent of Galactic BH transients in the thermal state (McClintock \& Remillard 2005), and are not seen in NS LMXBs of comparable luminosity.  The corresponding bolometric luminosities found from the spectral model are $\sim1.4\times10^{38}{\rm erg~s^{-1}}$, which equates to $\sim10\%$ Eddington for a 10 $M_\odot$ BH; consistent with known BH systems in the thermal state.  A NS primary emitting at ~${  L_{Edd}}$ seems unlikely based on comparison with Galactic analogues.  None of the brighter persistent (Z-track) sources exhibit spectral states as soft, and the only bright NS LMXB transient, XTE J1701-462, exhibited `Atoll-like' behavior (harder spectra) in the 60 days prior to its return to quiescence.   Spectral fits of absorbed disc blackbodies to M 31 XBs have also produced ${  kT_{in}}$ of 0.3-0.6 keV, such as XMMU J004144.7+411110 (Trudolyubov et al. 2006).  This source has an estimated unabsorbed luminosity of ${  3-4\times10^{37}}$~${\rm erg~s^{-1}}$ for the best-fit spectral model of an absorbed disc blackbody with ${  kT_{in}\sim0.6-0.8}$~keV.  Power law fits resulted in $  \Gamma\sim2.8-3.3$ and a column density far in excess of Galactic absorption, while the disc blackbody fits resulted in ${  N_H}$ only double that of the Galactic value, very similar behavior to our transient.   XMMU J004144.7+411110 also lacks an optical counterpart (${ m_v > 21}$).

CXOU J132527.6-430023  shows that our study of XB outside the Local Group is starting to probe lower-luminosity BH LMXB with properties more typical of Galactic sources (Fig. \ref{fig:lxkt}).  For the time being, Cen~A contains the only population of such objects ($10^{37}-10^{38}$~${\rm erg~s^{-1}}$, non-Local Group) that can be subjected to meaningful spectral fitting.  In future papers (Burke et al., in prep.) we will report on detailed spectral analysis of the Cen~A XB population of sources with $L_X$ $>10^{37}$~${\rm erg~s^{-1}}$ which will, for the first time, allow contrast between the nature of typical LMXBs in our galaxy with analogs from beyond the Local Group. 

\section{Conclusion}
\label{s:conc}

We believe that the following evidence makes a strong case for CXOU J132527.6-430023 being a transient BH LMXB candidate.

(i) Long-term Variability.  The absorbed flux varies by a factor $>100$ between quiescent and active states on timescales of months.  This behavior favors a low mass companion, as all dynamically confirmed BH HMXB systems (Cyg X-1 in the Milky Way, LMC X-1 \& LMC X-3 in the Large Magellanic Cloud) show persistent X-ray emission (McClintock \& Remillard 2005).  This stated, there is now a very strong case for a BH in HMXBs Cyg X-3 (Shrader et al. 2010) and SS433 (e.g. Blundell et al. 2008), both of which show considerable long-term variability. To date, such variability has been observed in very few NS LMXB systems, while all confirmed BH LMXB, confirmed by direct measurement of the radial velocity and spectral type of the companion, are transients.

(ii) Spectral fitting results.  The best-fit model of an absorbed disc blackbody with a measured inner disc temperature of $\sim 0.6$ keV is softer than is typical of NS LMXBs emitting at similar luminosities, and is characteristic of several observed BH systems in the thermal state.  The values of ${N_H}$ retrieved from fitting are consistent with those derived from source position in K-band optical depth maps that were presented in Kainulainen et al. (2009). This is not the case for the absorbed power law model, which required an extremely steep spectral slope, a much larger value of absorption column and achieved a less likely best-fit.

The parameter values of the best-fit absorbed disc blackbody model are typical of an accreting $10 M_\odot$ black hole in the thermally
dominant state. If indeed this source is an accreting BH, it is the first confirmed transient non-ULX black hole system in outburst to
be identified outside the Local Group.\\

This work was supported by NASA grant NAS8-03060. MJB thanks the STFC and the University of Birmingham for financial support.  RV is supported by NWO Vidi grant 016.093.305.  GRS acknowledges the support of an NSERC Discovery Grant. We also thank Jeff McClintock, Mike Garcia, Ewan O'Sullivan, Trevor Ponman \& Alastair Sanderson for useful  discussions.  Finally, we thank the anonymous referee for their very helpful comments.



\clearpage

\clearpage

\end{document}